\documentstyle[epsf]{aipproc}

\newif\ifpreprint
\preprinttrue       

\begin{document}

\ifpreprint \rightline{\bf UASLP--IF--99--01} \fi
\title{Experimental Techniques\ifpreprint\thanks{Course given at the
VII Mexican Workshop on Particles and Fields, Merida, Yucat\'an, Mexico,
November 10-17, 1999. Proceedings to be published by AIP.}\fi}
\author{J\"urgen Engelfried}
\address{Institito de F\'{\i}sica,
Universidad Aut\'onoma de San Luis Potos\'{\i}, \\
\'Alvaro Obreg\'on 64, Zona Centro, 78000 San Luis Potos\'{\i}, M\'exico\\
jurgen@ifisica.uaslp.mx, http://www.ifisica.uaslp.mx/\~\,$\!$jurgen}
\maketitle
\begin{abstract}
In this course we will give examples for experimental techniques
used in particle physics experiments.  After a short introduction,
we will discuss applications in silicon microstrip detectors, 
wire chambers, and single
photon detection in Ring Imaging Cherenkov (RICH) counters. A short discussion
of the relevant physics processes, mainly different forms of energy loss in 
matter, is enclosed.
\end{abstract}

\ifpreprint \setcounter{footnote}{0} \fi

\section*{Introduction}

In this course we will not try to reproduce standard text books about
detectors (see for example~\cite{jurgen:fernow,jurgen:kleinknecht}) and
descriptions of interaction with matter (a good summary can be found
in~\cite{jurgen:pdg}), 
covering in great details all aspects of experimental high energy physics.
Due to the time restrictions ($3\times 50\,\mbox{min}$ were assigned by
the organizers for this course) we will rather discuss examples on the
use of some detector families.  The selection is highly biased, since
the author decided to use examples he knows best, e.g.\ he either worked on
some of the detectors directly, or they were part of an experiment he
participated in.

In a particle physics experiments detectors of various kinds are placed
downstream (some even upstream) of the fixed target or surrounding
a collision point of colliding beams. In general different detectors and 
electronic apparatus
are required to perform the different tasks most experiment require:
tracking, momentum analysis, particle identification, neutral particle 
detection, triggering, and data acquisition.  Clearly a very important
part is also the analysis of data, which comes in two parts:  the
event reconstruction, using all the detector information available (which 
requires for example a good alignment and calibration of all parts), and
the final physics analysis.  All these pieces cannot live by them self,
to perform a successful experiment requires that all components, be it
hardware or software, work all together to reach the final goal:  A good
and significant physics result.
\newpage
\ifpreprint\pagestyle{plain}\fi

Usually resources (both money and person-power) are limited when designing,
building and operating an experiment.  Careful consideration is necessary
to decide, if some fancy or expensive detector is really necessary to
obtain the physics goals or if a simpler, less expensive version would be
sufficient, and the free resources can be applied to more essential parts
of the experiments.  Sometimes there are also political constraints that
make a clear technical decision more complicated.  All these arguments
hold for the hardware as well as for the software part of the experiment.   

In the following we will discuss generically MWPCs, 
the properties of a silicon microstrip
array used in the SELEX experiment, 
and a longer part about RICH detectors used in WA89, SELEX, and CKM.

\section*{Multiwire Proportional and Drift Chambers}

This section should be seen as an introduction to energy loss in matter
and its use for particle detection, with these devices as examples.

Long time ago Geiger invented his counter:  A charged particle will ionize
gas, and the liberated electrons drift under the influence of an electric
field to a thin counting wire.  Close to the wire the field is strong enough
to accelerate the electrons to ionize again further gas atoms or molecules
(gas multiplication).
In typical applications multiplication gains of $10^5$ or more can be obtained.
The electrons drift to the wire, producing there a fast signal which is
usually not resolved by the preamplifier, but the backwards drifting 
ions induce an additional, slower, signal, which can be easily detected.
 
\looseness=-1
Without any further consideration this will actually lead to a spark,
usually something not welcome (but in the 60's ``spark chambers'' where used
for particle detection), since in the multiplication avalanche not only 
ions, but also excited atoms or molecules are produced, which will de-excite
with the emission of a (UV)-photon, which can again ionize.  Two tricks already
used by Geiger help to avoid this: 1)~A sufficiently large resistor is included
in the HV line, so that the current drawn will lead to a voltage drop. 
2)~The addition of so-called quencher gases, usually some alcohol, to the 
detector gas, who will absorb the UV-photons without being ionized.

In the 60's several groups, most famous the group around
Charpak at CERN, started to develop counters later known as MWPC: Instead
of one wire, a lot of wires are stretch parallel, with the electrons 
drifting to the {\sl nearest} wire.  This allows construction of larger
area detectors, something necessary in the time when people tried to develop
electronic detectors to replace the bubble chambers.
The resolution of the detector is given by the wire distance, and space
points can be obtained be putting MWPCs under different angles.
In practice the wire distance is limited to about $1\,\mbox{mm}$ in small
size chambers, and even more in larger areas for two reasons:  The wires
have to be stretched, supported by a strong frame, and, even stretched, 
electrostatic deflection has to be taken into account.

\looseness=-1
In the mid-60's, a new idea came up, first realized in Heidelberg by
Heintze and Walenta~\cite{jurgen:Heintze}:  
If the time a particle passes the counting gas is known,
e.g.\ by using for example an scintillator somewhere in the experiment, 
the drift time of the electrons from the point of ionization to the wire
contains also space information. Putting the wires further apart, and
forming with cathodes a homogeneous electric field (with exception close
to the wire), a constant drift velocity in the order of several
$\mbox{cm/}\mu\mbox{sec}$ is observed. The resolution, limited by
the diffusion of the electrons, obtained with
drift chambers can be well below $100\,\mu\mbox{m}$, even with drifts
of 10's of~cm.  The advantage is clearly the reduced number of wires and
readout channels, with the additional cost of the need of measuring the
drift time with some sort of TDC.

The energy loss of moderately charged particles (other than electrons)
in matter is primarily ionization.  The mean rate of energy loss is given
by the Bethe-Bloch equation:
\begin{equation}
-{{dE}\over{dx}} = K z^2 {{Z}\over{A}} {{1}\over{\beta^2}}\left[
{{1}\over{2}}\ln{{2 m_e c^2 \beta^2 \gamma^2 T_{\rm max}}\over{I^2}}
- \beta^2 - {{\delta}\over{2}}\right] \label{jurgen:bethe}
\end{equation}
Here $K$ is some constant, $ze$ the charge of the particle, $Z$ and $A$ 
the atomic number and charge of the medium, $T_{\rm max}$ the maximum 
energy of a free electron after one collision, $I$ the mean excitation
energy of the medium, and $\delta$ is a correction factor; 
the other symbols have there usual obvious meaning. $dx$ is measured here in
units of $\mbox{g}/\,\mbox{cm}^2$.
The Bethe-Bloch formula only describes the mean energy loss; for finite
path lengths, there are significant fluctuations in the actual energy loss.
The distribution is skewed towards high values, described by the
Landau distribution. Only for a thick layer the distribution is nearly
Gaussian.
 
As seen from eq.~\ref{jurgen:bethe}, for $\beta\gamma\gtrsim 3$ the energy
loss has a so-called ``relativistic raise''. If the momentum of the particle
is known, this can be used to identify the particle.  The problem is that
one has to sample the energy loss several times to be able to extract the
average loss (Landau!).  This is explicitly done in so-called
``jet chambers''.  An most up-to-date example is the OPAL central jet
chamber~\cite{jurgen:opala,jurgen:opalb,jurgen:opalc}, 
where a normal track gets measured at 159~points, using 
$4\,\mbox{m}$ long wires spaced by $1\,\mbox{cm}$.  
The 3-dim space information ($r$:\ wire number, $\phi$: drift time,
$z$: charge division)
gets used to measure
the momentum (a magnetic field is present) and the total charge information
helps to identify the particle.
Another application is a TPC (Time Projection Chamber); the whole drift
volume contains only gas with parallel electrical and magnetic field 
(to reduce the diffusion), and the electrons drift to the end plates, where
wire or pads are used to obtain space information.

\section*{Silicon Microstrip Detectors}

In the 1980's, silicon microstrip detectors became used heavily in HEP.
They are absolutely necessary to measure properties of particles
containing charm and beauty quarks.  Examples for very successful experiments
using this kind of detectors include E691 at Fermilab, WA82 at CERN,
and, in colliders, CDF, the 4 LEP experiments (Aleph, DELPHI, L3, OPAL), and
the HERA experiments.  Today there are a lot of experiments using
silicon microstrips, with channel counts up to 1~million or more.

\begin{figure}[htb]
\begin{center}
\leavevmode
\vspace{6cm}
\includegraphics{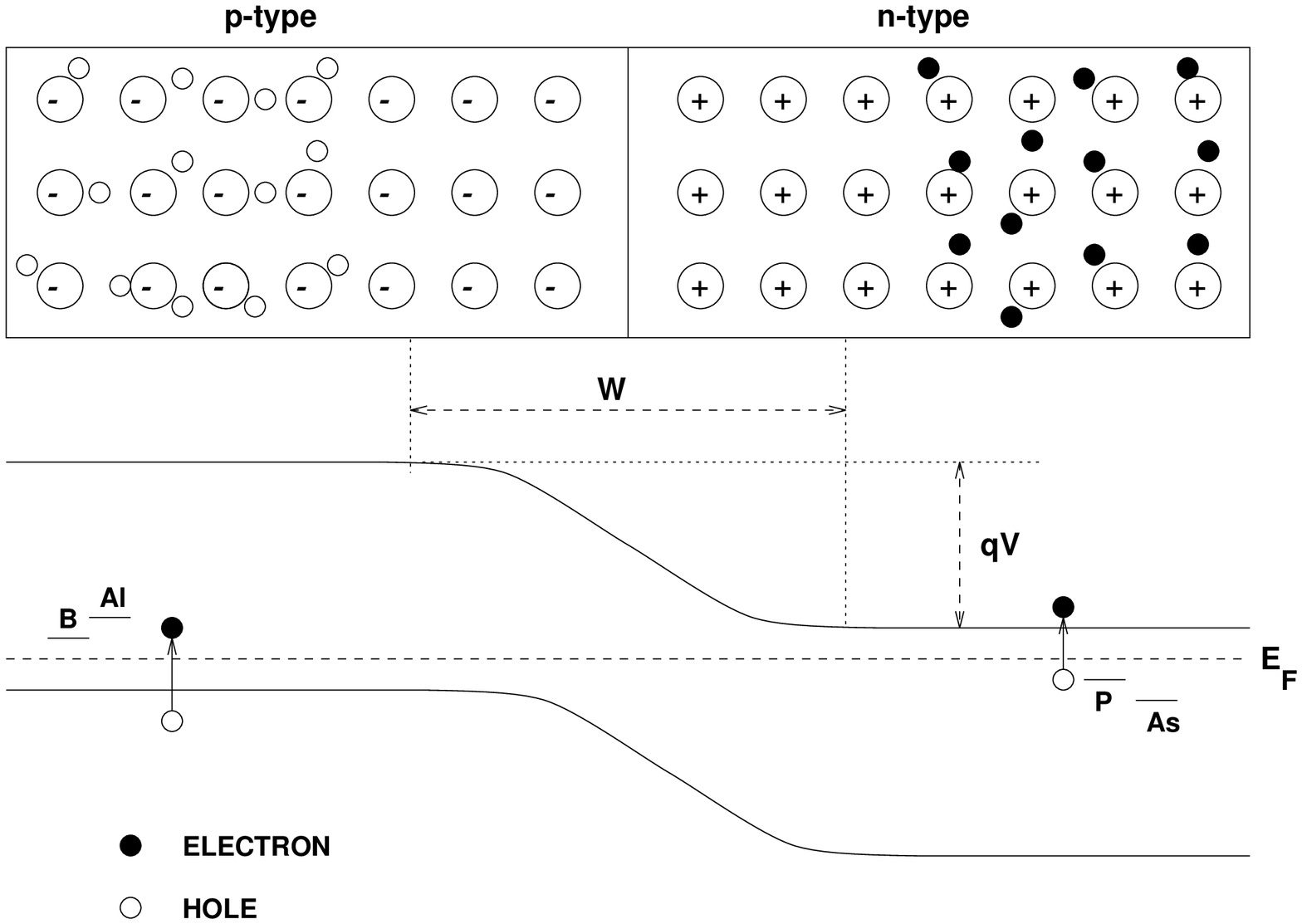}
\includegraphics{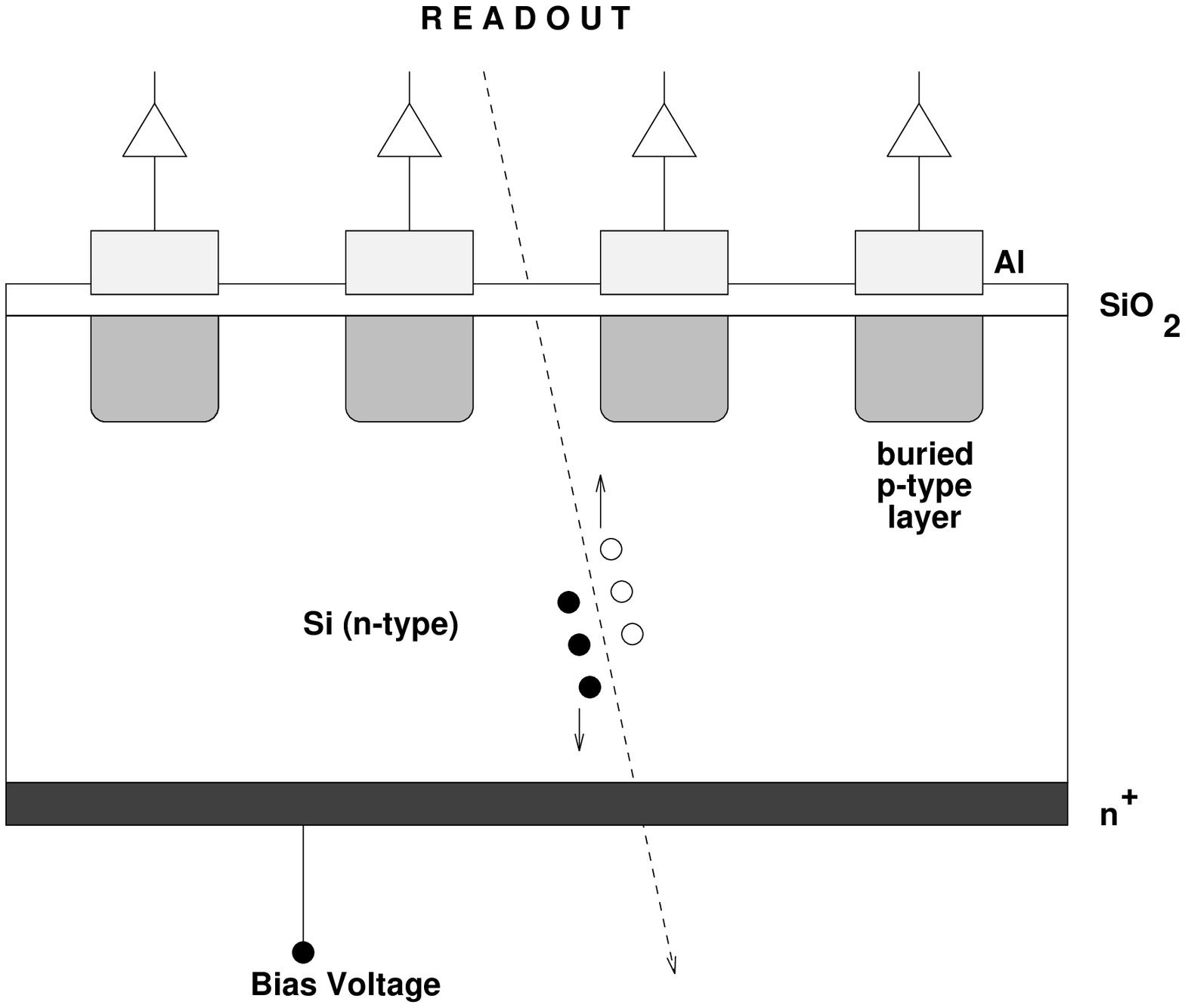}
\caption{Left: Model of a pn-junction in a semiconductor. Right: Schematic
drawing of silicon strip detector.
Both figures taken from~\protect\cite{jurgen:prakash}.}
\label{jurgen:pnjunction}
\end{center}
\end{figure}

\looseness=-1
The detector allows to measure with a precision of down to a few $\mu\mbox{m}$
the one-dimensional position of a passing charged track.  Newer devices,
the so-called pixel detectors, measure a two-dimensional position.
The detector uses as basic detection device a pn-junction, shown in 
fig.~\ref{jurgen:pnjunction} left,
a diode which is operated in blocking direction with a sufficiently high 
voltage so that the entire device is depleted, e.g.\ there are no free
electron-hole pairs (also called charge carriers).  In one device several
(up to several thousand) of these pn-junctions are operated, arranged
in parallel strips.  Should a charged particle pass through the detector
(see fig.~\ref{jurgen:pnjunction}~right), new electron-hole pair are created
and one of the carrier types will drift towards the nearest strip. In 
Silicon, the energy loss $dE/dx\approx 3.8\,\mbox{MeV/cm}$, and the energy 
needed to create one electron-hole pair is 
$3.6\,\mbox{eV}$\footnote{The band gap in Silicon is only $1.1\,\mbox{eV}$,
but Silicon is an indirect semiconductor.}, so in a
typically $300\,\mu\mbox{m}$ thick detector about $3\cdot10^4$ pairs will be
created. 

\looseness=-1
The construction of the detector itself seems to be under control today.
There are several companies available which will produce the silicon
detector with a well understood process.  The smallest strip distance
used today is $10\,\mu\mbox{m}$, so that the structure is actually much
simpler than the achieved sub-micron structures in todays semiconductor chips.
The real challenge in these detectors is the readout:  Imagine a 
$5\,\mbox{cm}\times5\,\mbox{cm}$ detector with $10\,\mu\mbox{m}$ strip 
distance: 5000 strips with there small signals have to be readout. Every
single strip needs a preamplifier, and some kind of signal detection like
a discriminator, otherwise noise will overwhelm the data acquisition.
To reduce the number of cables (anyway, how to have a cable 
every $10\,\mu\mbox{m}$?)\ it would be nice to chain several channels 
together, at best even all 5000.  The chips should then be clever enough only
to send a strip number to the data acquisition, e.g.\ the signal gets 
digitized and zero suppressed already at the detector.

\begin{figure}[htb]
\begin{center}
\leavevmode
\vspace{6cm}
\includegraphics{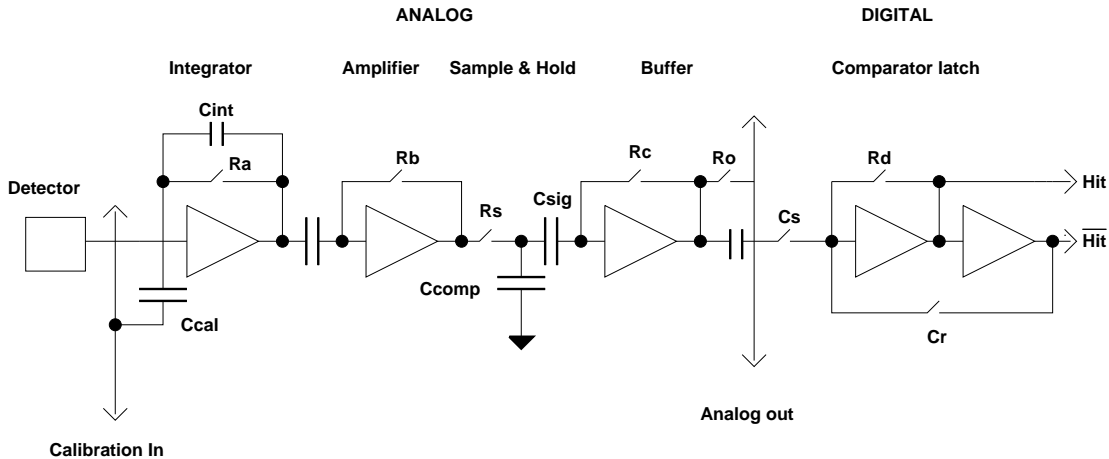}
\caption{Schematic of a single SVX channel.  
One SVX1 chips contains 128~channels.
Figure taken from~\protect\cite{jurgen:prakash}.}
\label{jurgen:svx}
\end{center}
\end{figure}

\looseness=-1
A system like this, called SVX~\cite{jurgen:lblsvx}, 
was developed about 10~years ago by LBL for collider experiments (CDF),
and also used 
in WA89~\cite{jurgen:wa89silicon} and 
SELEX~\cite{jurgen:prakash,jurgen:selexsilicon}.  
The basic layout of the SVX system is shown in fig.~\ref{jurgen:svx}.
The current is integrated onto a capacitor as long as $R_a$ is closed. The
charge is then transfered via a sequence of switch operations and compared
to a pre-stored charge.  If the signal charge is bigger, the channel offers
its number to the readout.  Up to 64~chips with 128~channels each can be
chained for readout.  Since this chip was developed for collider, the fact
that the chip is integrating is not very important, since a clear cycle can
be performed before every collision.  In fixed target operation it is not 
known when an interaction will happen, so in general the chip will integrate
several interactions until a clear cycle is performed, which has to be
closely coupled with the trigger of the experiment since during
a clear the detector is not sensitive.  Depending on the beam rate, the
number of interactions and the sensitivity of the experiment to out-of-time
tracks, the
ratio of integration and clear time has to be optimized for the experiment.

\begin{figure}[htb]
\begin{center}
\leavevmode
\vspace{9.5cm}
\includegraphics{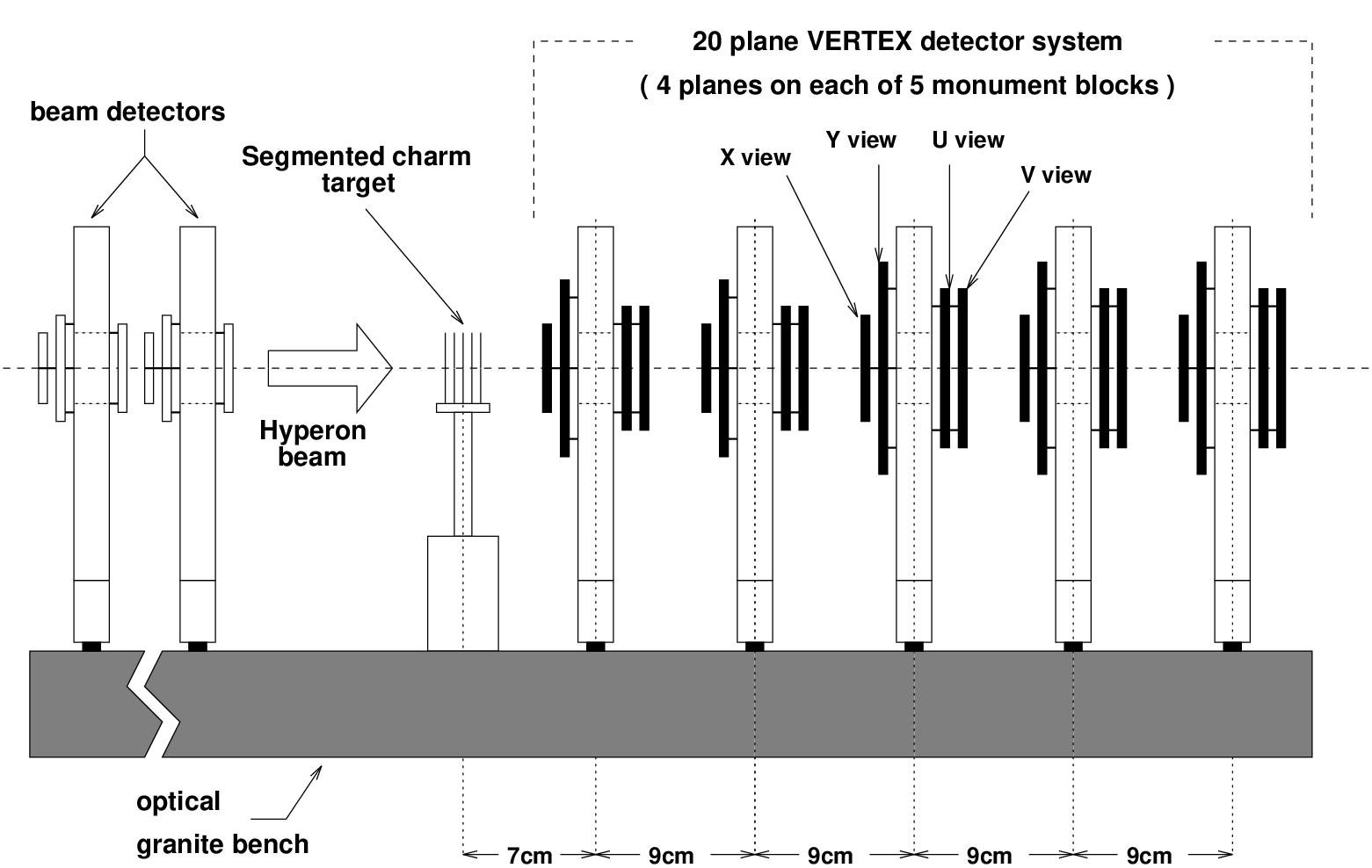}
\caption{Layout of the SELEX vertex detector. After the target is a total
of 20~planes with $20\,\mu\mbox{m}$ and $25\,\mu\mbox{m}$ 
strip distance in 4 orientations.
Figure taken from~\protect\cite{jurgen:prakash}.}
\label{jurgen:siliconlayout}
\vspace{9.25cm}
\includegraphics{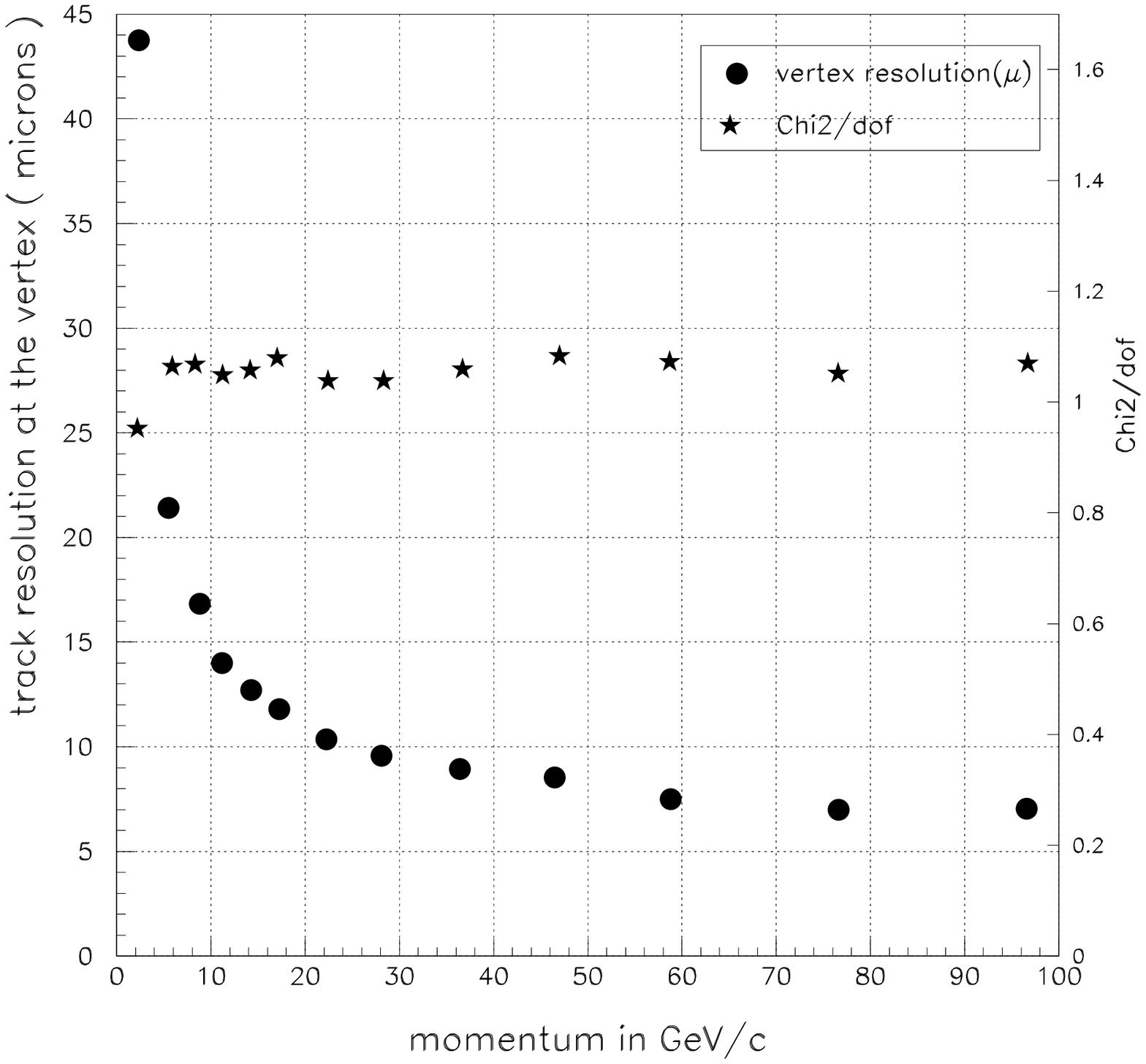}
\caption{Mean $\chi^2/\mbox{dof}$ and vertex track resolution as a function
of momenta. Figure taken from~\protect\cite{jurgen:prakash}.}
\label{jurgen:silresol}
\end{center}
\end{figure}

\looseness=-1
The layout of a typical fixed target vertex detector is shown in 
fig.~\ref{jurgen:siliconlayout}.
Tracks originating from the targets are transversing the silicon planes 
oriented in 4 different orientations (rotated by $45^\circ$) to allow
the reconstruction of tracks in space.  They eventually get fit to form
a vertex, and the obtained resolution is shown in
fig.~\ref{jurgen:silresol}.
At high momentum the resolution is limited by the strip distance, but at
lower momentum multiple scattering becomes more and more important.
Nevertheless, the fit takes all error contributions correctly into account,
as seen from a constant $\chi^2=1$ for all momenta.  This is another lesson
to learn:  more detectors is not always good.

Another nice example is the silicon drift detector.  The device developed
for ALICE was presented in this workshop by E.~Crescio~\cite{jurgen:sildrift}.

\clearpage
\section*{Ring Imaging Cherenkov Counters}

\subsection*{Introduction}

Even though the basic idea of determining the velocity of charged particles
via measuring the Cherenkov angle was proposed in 1960~\cite{jurgen:ArtR},
and in 1977 a first prototype was successfully operated~\cite{jurgen:desrich},
it was only during the last decade that Ring Imaging Cherenkov (RICH) 
Detectors 
were successfully used in experiments.  A very useful collection of 
review articles and detailed descriptions can be found in the proceedings
of three international workshops on this type of
detectors, which were held in 
1993 (Bari, Italy)~\cite{jurgen:bari}, 
1995 (Uppsala, Sweden)~\cite{jurgen:upsalla}, 
and 1998 (Ein Gedi, Israel)~\cite{jurgen:israelproc}, respectively.

Charged particles with a velocity $v$ larger than the speed of light 
in a medium with refractive index $n$ will emit Cherenkov radiation under an
angle $\theta$, given by~\cite{jurgen:cherenkov}
\begin{equation}
\cos\theta = {{1}\over{\beta n}} \label{jurgen:cherenkovangle}
\end{equation}
with $\beta=v/c$, $c$ being the speed of light in vacuum.
The number of photons $N$ emitted per energy interval $dE$ and path length $dl$
is given by~\cite{jurgen:franck}
\begin{equation}
{{d^2N}\over{dE dl}} = {{\alpha}\over{\hbar c}}
\left(1 - {{1}\over{(\beta n)^2}}\right)
= {{\alpha}\over{\hbar c}} \sin^2\theta
\end{equation}
or, expressed for a wavelength interval $d\lambda$,
\begin{equation}
{{d^2N}\over{d\lambda dl}}
= {{2 \pi \alpha}\over{\lambda^2}} \sin^2\theta \label{jurgen:dndl}
\end{equation}

By measuring the Cherenkov angle $\theta$ one can in principle
determine the velocity
of the particle, which will, together with the momentum $p$ obtained via
a magnetic spectrometer, lead to the determination of the mass and therefor
to the identification of the 
particle\footnote{For this reason Cherenkov detectors
are usually described under the chapter ``particle identification'' in 
particle detectors books.}.

\begin{figure}[htb]
\begin{center}
\leavevmode
\epsfysize=10cm
\epsfbox{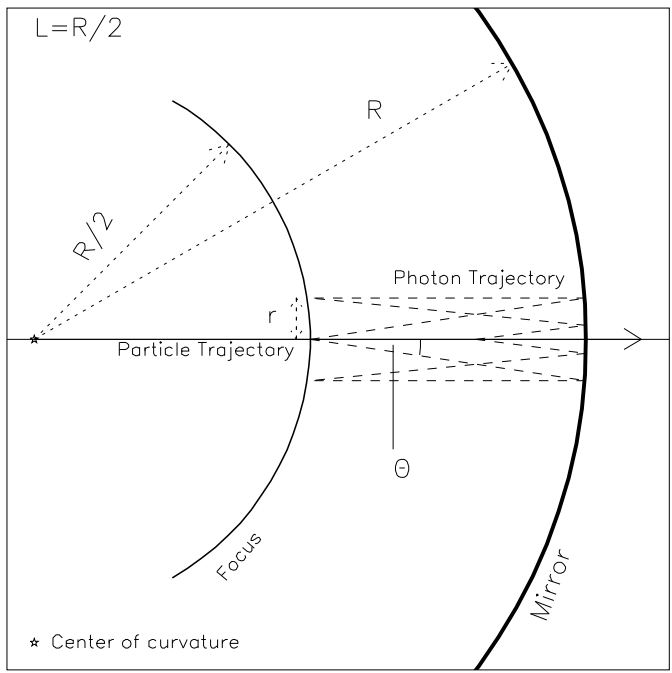}
\caption{Schematic of a RICH detector}
\label{jurgen:richprinzipal}
\end{center}
\end{figure}

\begin{figure}[htb]
\begin{center}
\leavevmode
\epsfxsize=\hsize
\epsfbox{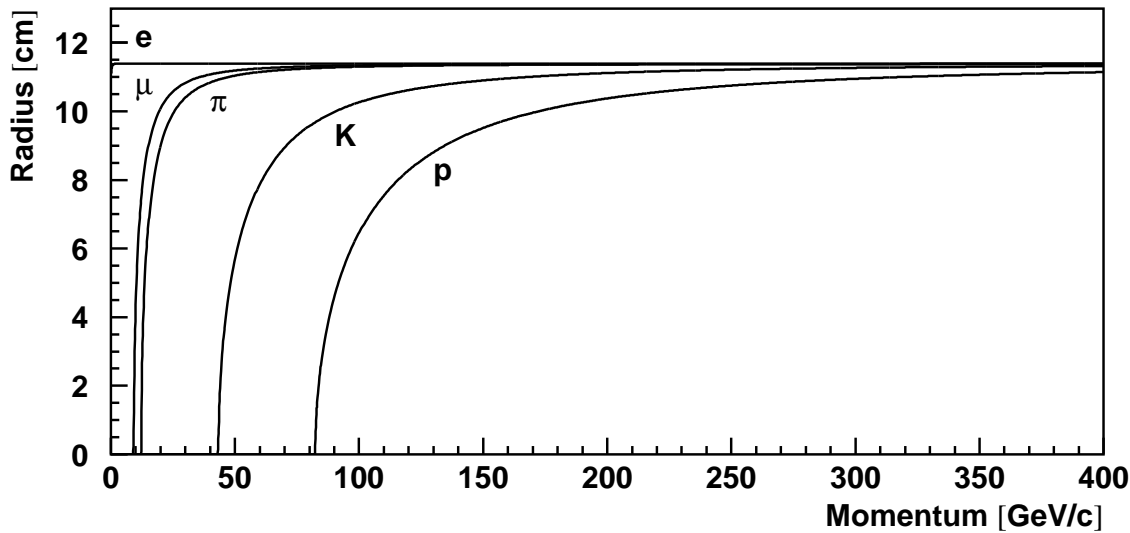}
\caption{Ring radii for different particles as a function of momentum,
in the
case of the SELEX RICH detector ($R=1982\,\mbox{cm}$, radiator Neon).} 
\label{jurgen:radii}
\end{center}
\end{figure}

Neglecting multiple scattering and energy loss in the medium, all the Cherenkov
light (in one plane) is parallel, and can therefor be focused (for small
$\theta$) with a spherical mirror (radius $R$) onto a point, as shown in 
fig.~\ref{jurgen:richprinzipal}. 
Since the emission is symmetrical in the azimuthal angle around the
particle trajectory, this leads to a ring of radius $r$ in the
focus, which is itself a sphere with radius $R/2$. The  radius $r$ is given by 
\begin{equation}
r = {{R}\over{2}} \tan\theta
\end{equation}
The dependence of the ring radius on the momentum for different particles
is shown in fig.~\ref{jurgen:radii}

\begin{figure}[htb]
\begin{center}
\leavevmode
\epsfxsize=\hsize
\epsfbox{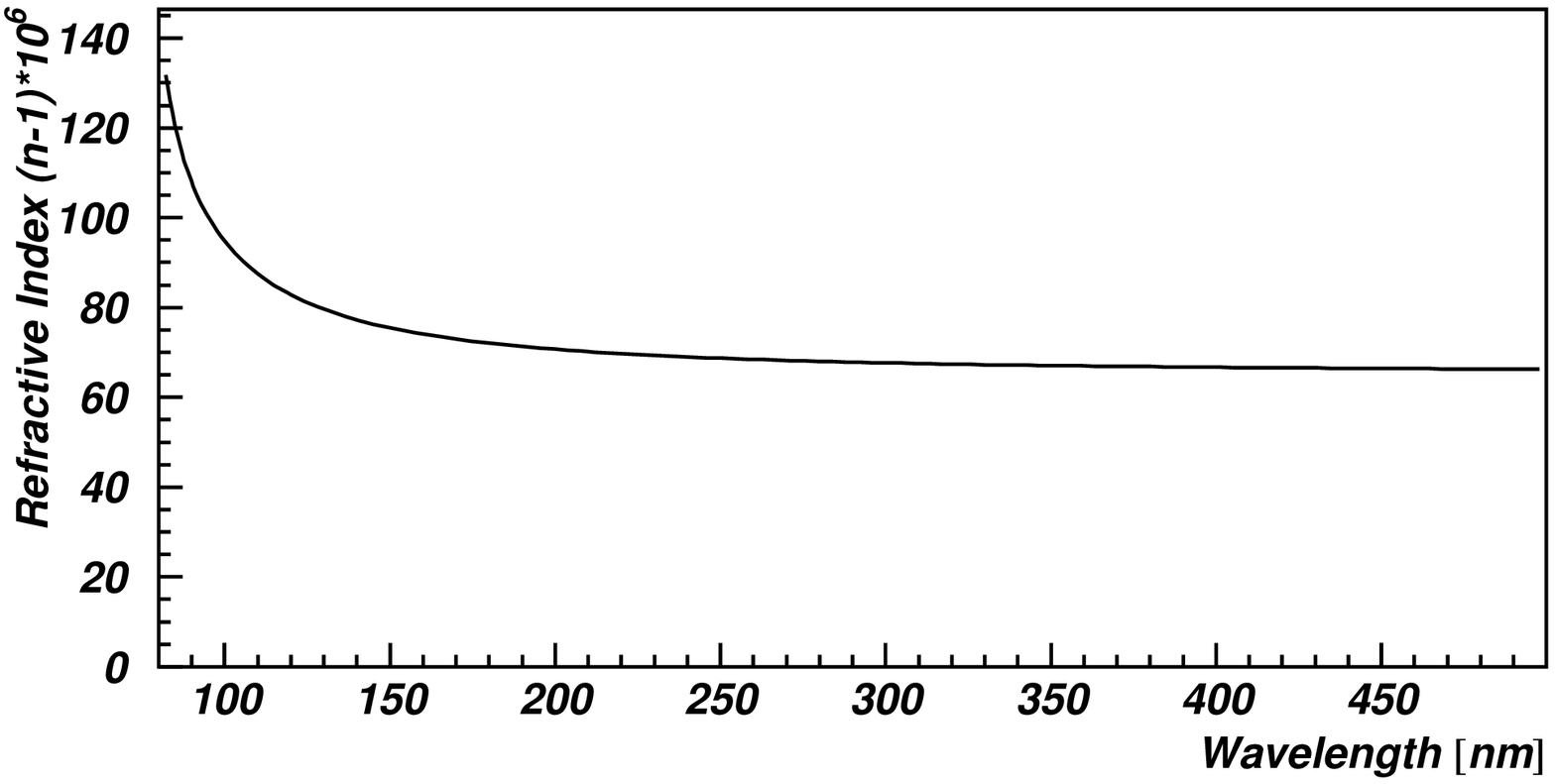}
\caption{Refractive index $n-1$ for Neon as a function of 
wavelength at STP~\protect\cite{jurgen:abjean}.}
\label{jurgen:neon}
\end{center}
\end{figure}

Since the number of photons is $\propto\lambda^{-2}$, most of the light is
emitted in the VUV range.  To fulfill equation~\ref{jurgen:cherenkovangle},
the refractive index has to be $n>1$, so there will be no Cherenkov radiation
in the x-ray region.  Also it is very important to remember that $n$ is a 
function of the wavelength ($n=n(\lambda)$, chromatic dispersion)
and most materials have a
absorption line in the VUV region, where $n\to\infty$,
as shown in fig.~\ref{jurgen:neon}, using Neon as example.
Since usually the
wavelength of the emitted photon is not measured, this leads to a smearing
of the measured ring radius, and one has to match carefully the wavelength
ranges which one wishes to use:  Lower wavelengths gives more photons, but
larger chromatic dispersion.

A very useful formula is obtained by integrating eq.~\ref{jurgen:dndl}
over $\lambda$ (or $E$),
taking into account all efficiencies etc., obtaining a formula for
the number of detected photons $N_{\rm ph}$~\cite{jurgen:desrich}:
\begin{equation}
N_{\rm ph} = N_0 L \sin^2\theta
\end{equation}
where $N_0$ is an overall performance measure (quality factor) of the detector,
containing all the details (sensitive wavelength range, efficiencies), and
$L$ is the path length of the particle within the radiator. A `` very good''
RICH detector has $N_0=100\,\mbox{cm}^{-1}$, which gives typically around
$10$ to $15$ detected photons ($N_{\rm ph}$) per $\beta=1$ ring.

The usual construction of a RICH detector is to use a radiator length of
$L= R/2$, e.g.\ equal to the focal length; but any other configuration, like
folding the light path with additional (flat) mirrors is possible.

All the presented arguments and the drawing in fig.~\ref{jurgen:richprinzipal}
only work for small $\theta$, which is always fulfilled in gases,
since $n$ only differs little from $1$.  Also important is the fact, that,
should the particles not pass through 
the common center of curvature of mirror(s)
and focal spheres, the ring gets deformed to an ellipse or, in more extreme
cases, to a hyperbola. If the photon detector is able to resolve this, and the
resolution is needed for the measurement, these deviations from a perfect
circle have to be taken into account in determining the velocity $\beta$.
In general this effect can be  neglected, and all parallel particles (with
the same $\beta$) will give the same ring in the focal surface, due to the
fact that all emitted Cherenkov light is parallel.  The position of the 
ring center is determined by the angle of the tracks, not by their position.

In the following, we will describe two RICH detectors used in experiments,
and a new application for RICH detectors for a new, proposed experiment.
The author works or worked on all of them, so the selection is clearly biased.
Even so, we feel that they are good examples for the use of this kind
 of detectors.

\newpage
\subsection*{The CERN Omega-RICH}

In the middle of the 1980's, first attempts were made to apply the
prototype results obtained by S\'eguinot and Ypsilantis~\cite{jurgen:desrich}
to experiments in a larger scale.  One of these attempts was performed
at the CERN Omega facility in the West Hall. Experiments WA69 and WA82
tried to use this detector for there analysis, but only succeeded partly.
An overview about this history can be found in~\cite{jurgen:wa89bari}.
When in 1987 a new experiment, later named WA89, 
was proposed~\cite{jurgen:wa89loi,jurgen:wa89proposal}, an
important part was a necessary upgrade of this detector for the use by
this new experiment.  Two main parts where changed:  New photon detectors
using TMAE as photo sensitive component, and new mirrors to perform the
focusing.  Details about the detector 
can be found 
in~\cite{jurgen:wa89bari,jurgen:doktorarbeit,jurgen:wa89wire,%
jurgen:wa89upsalla,jurgen:wa89israel}.

\begin{figure}[htb]
\begin{center}
\leavevmode
\epsfysize=8.5cm
\epsfbox{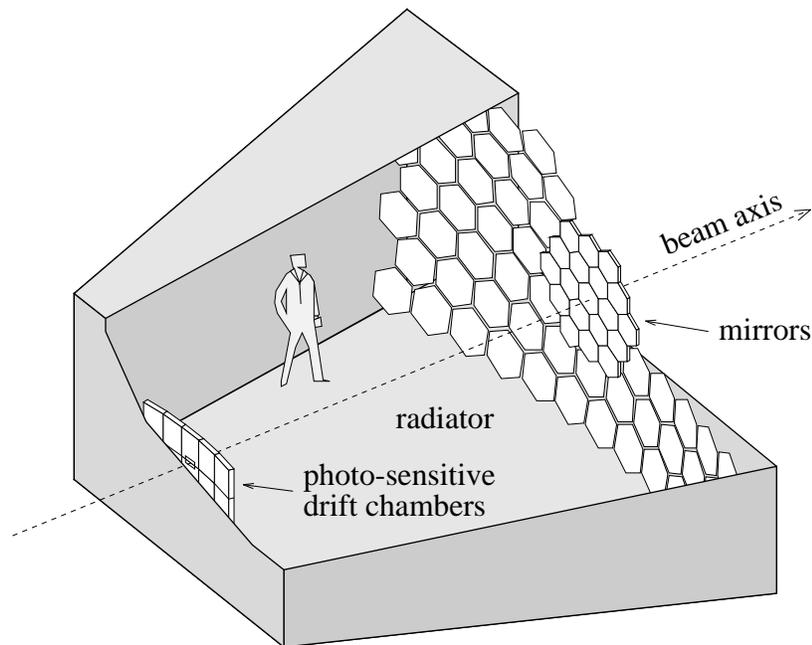}
\caption{Overall view of the CERN Omega RICH detector, used in experiment
WA89.  The radiator gas is pure Nitrogen, so usually there is no person
within the radiator box.} 
\label{jurgen:wa89box}
\end{center}
\end{figure}

As seen in the overall layout of the detector (fig.~\ref{jurgen:wa89box}), a
RICH detector is basically a simple device:  a big box, some mirrors at the
end, and photon-sensitive detectors at the entrance.
The real challenge is to combine all the parameters together to obtain 
a perfect match for the overall system.

\looseness=1
The size of the radiator box 
and the photon detector is given by the angular distribution of tracks which
have to be identified at the location of the detector.  Since usually this
detector is placed behind a magnetic spectrometer, and the momentum spectrum
of the interesting tracks depends on the physics goals of the experiment,
the surfaces to cover have to be determined for every setup and experiment,
usually with Monte Carlo simulations during the design phase of the
experiment~\cite{jurgen:doktorarbeit}.
In the case of WA89, the mirror surface needed was
about $1\,\mbox{m}\times 1.5\,\mbox{m}$, much smaller than the 
$4\,\mbox{m} \times 6\,\mbox{m}$ covered by the original Omega-RICH. It was
therefor decided to replace only the central mirrors with smaller 
(as seen in fig.~\ref{jurgen:wa89box}),
higher surface quality mirrors to obtain better resolution.

\begin{figure}[htb]
\begin{center}
\leavevmode
\epsfysize=10cm
\epsfbox{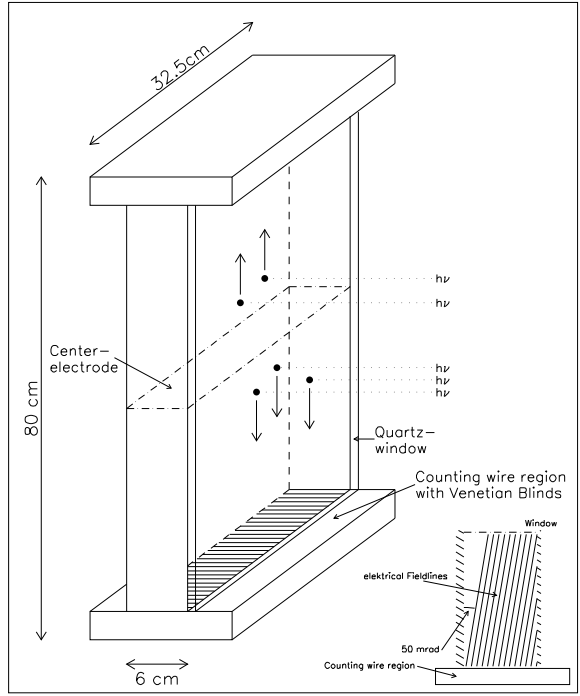}
\caption{Schematic of the Omega-RICH photon detector module.
$h\nu$ denotes Cherenkov photons. At the central electrode a voltage
of $-40\,\mbox{kV}$ was applied.
The inset shows a side view, demonstrating the electrical 
field lines (figure taken from~\protect\cite{jurgen:doktorarbeit}).} 
\label{jurgen:wa89chamber}
\end{center}
\end{figure}

\begin{figure}[htb]
\begin{center}
\leavevmode
\epsfxsize=\hsize
\epsfbox{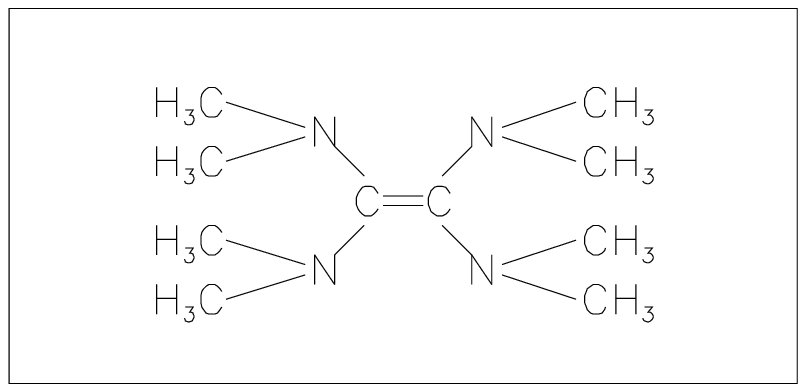}
\caption{Structural formula for Tetrakis(dimethylamino)-ethylen (TMAE).
This derivate of Ammonium is the molecule with the lowest ionization
energy ($5.3\,\mbox{eV}$).}
\label{jurgen:tmaestruc}
\leavevmode
\epsfysize=8.13cm
\epsfbox{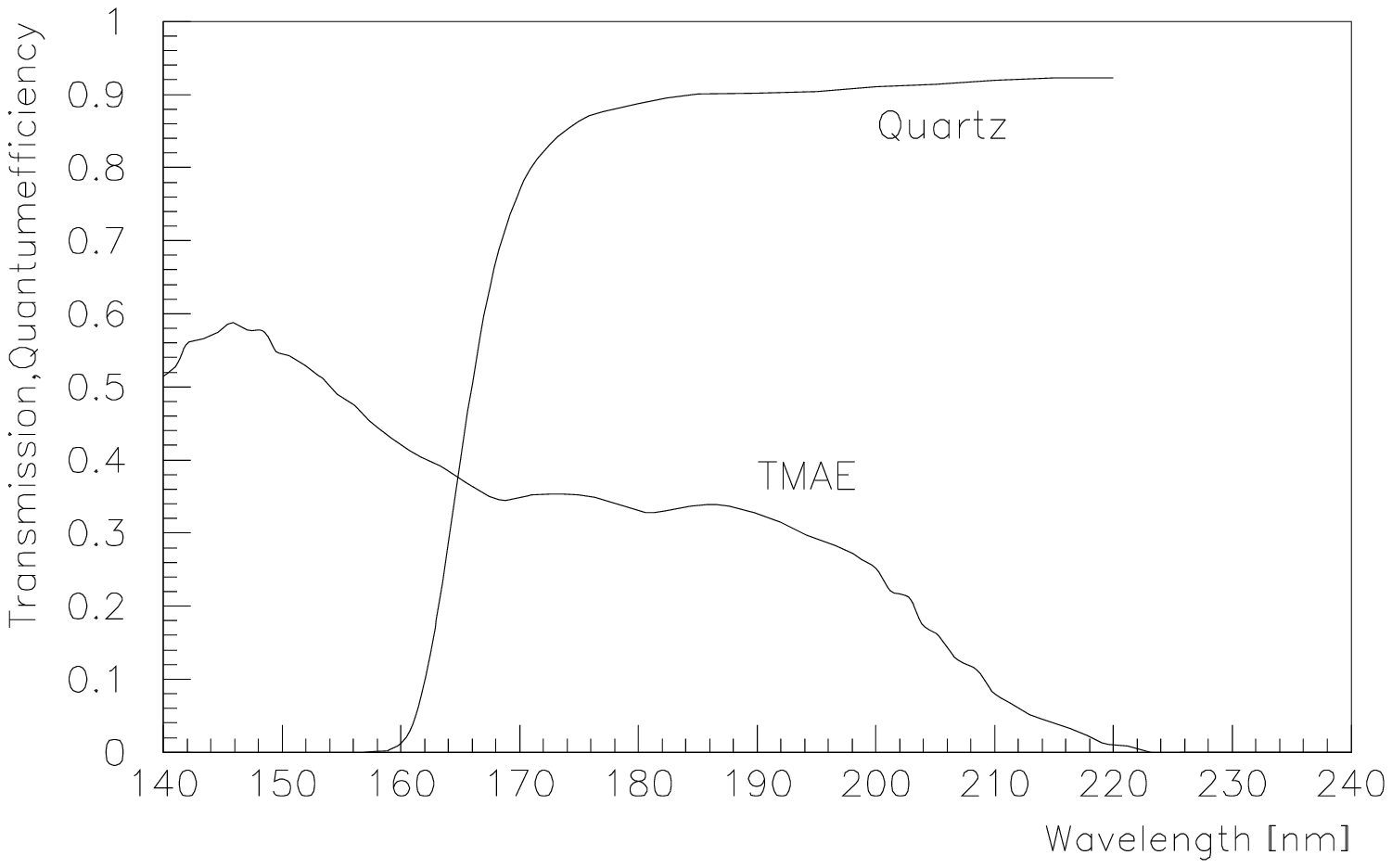}
\caption{Transmission of a quartz window actually used in the Omega-RICH,
together with the quantum efficiency (probability of releasing an electron
via photo-effect) of TMAE~\protect\cite{jurgen:holroyd}, 
both as a function of wavelength.} 
\label{jurgen:tmaeqe}
\end{center}
\end{figure}

The detector surface was calculated to be $1.6\,\mbox{m}\times0.8\,\mbox{m}$,
with a spatial resolution of a few millimeters for every detected photon.
The pixel size could therefor not be much bigger than also a few millimeters,
leading to about $100000$ pixels in the detector plane.  The solution
was to build drift chamber (TPC) modules, 
shown in fig.~\ref{jurgen:wa89chamber}, 
covering an area of
$35\,\mbox{cm}\times 80\,\mbox{cm}$, and approximating the focal sphere
with a polygon of 5~modules.  After passing a $3\,\mbox{mm}$ thick quartz 
window (to reduce absorption), the photons hit TMAE
(see fig.~\ref{jurgen:tmaestruc}) molecules, 
converting via photo-effect into an single electron. TMAE is present
with a concentration of about $0.1\,\mbox{\%}$ with the driftgas, which
is otherwise pure ethan. The use of a quartz window together with TMAE
as photo-sensitive gas leads that the detector is only sensitive in a small
wavelength range between $165\,\mbox{nm}< \lambda <230\,\mbox{nm}$, as
demonstrated in fig.~\ref{jurgen:tmaeqe}.
TMAE has a very low vapor pressure, so that at ambient temperatures the
molecule is saturated within a gas.  To obtain a short enough conversion
length of around $1\,\mbox{cm}$ (otherwise the conversion would occur
to far away from the focal plane and lead to an addition contribution to
the resolution), the drift gas (ethane) is led through a bubbler, containing
TMAE liquid at $30^\circ\,\mbox{C}$. 
This means that everything after the bubbler,
e.g.\ the whole detector including radiator box, had to be heated to
$40^\circ\,\mbox{C}$ to avoid condensation.  Other unpleasant properties
of TMAE include a high reactivity with Oxygen, producing highly 
electro-negative
oxides, which will attach an electron easily, change the drift velocity
by a factor of several thousand, leading to a loss of electrons.  Since
the signal is a single electron (photo-effect!)\ this is catastrophic.  The
counting gas had an Oxygen contents of $<1\,\mbox{ppm}$.

Mostly due to the presence of TMAE,  the operation of this detector was 
not trivial.  All parameters were monitored electronically, and
hardware limits on some critical
parameters (like temperature, Oxygen content of Ethane) lead to a automatic
shutdown of the detector, waiting for an expert to arrive in the experimental
hall.
  
Once the electron was released, it was drifting under the influence of an 
electric field of $1\,\mbox{kV/cm}$ 
(drift velocity $5.4\,\mbox{cm/}\mu\mbox{m}$) upwards or downwards 
over maximal $40\,\mbox{cm}$ towards
$6\,\mbox{cm}$ long counting wires (gold-coated tungsten, $15\,\mu\mbox{m}$
diameter), spaced by $2.54\,\mbox{mm}$. The two-dimensional 
spatial information about the
conversion point of the photon is obtained with the position of the wire and
the drift time of the electron. In total, 1280 wires were used in the
detector.  An additional complication was that the charged particles itself
where passing through the chambers, leaving a $dE/dx$ signal of several
hundred electrons, which is to be compared to the single electron which is
our signal.  This leads to increased requirements for the wire chambers
(sensitive, e.g.\ sufficient multiplication, to single electrons,
but no sparking with several hundred electrons) and to the preamplifier
electronic (not too much dead time after a big pulse).

The overall resolution allowed the separation of pions and kaons up to 
a momentum of about $100\,\mbox{GeV/}c$, which was exactly the design goal.
This lead to a good number of physics 
results~\cite{jurgen:wa94a,jurgen:wa89a,jurgen:wa89b,jurgen:wa89c,%
jurgen:wa94b,jurgen:wa89d,jurgen:wa89e,jurgen:wa89f,jurgen:wa89g,jurgen:wa89h},
which would not have been possible to obtain without the RICH detector.

\clearpage
\subsection*{The SELEX Phototube RICH Detector}

At Fermilab an new hyperon beam experiment, called SELEX, was proposed
in 1987~\cite{jurgen:selexproposal}.  The key elements to perform a successful
charmed-baryon experiments are 1)~a high resolution silicon vertex detector
and 2)~a extremely good particle identification system based on RICH.
During the following years, a prototype for the SELEX RICH was constructed and
tested successfully~\cite{jurgen:protos}, 
based in some part on experience gained
by our Russian collaborators~\cite{jurgen:sphinx,jurgen:sphinxnew}.
The real detector was constructed in 1993-1996, ready for the SELEX 
data taking period from July~1996 to September~1997.
First results for the final detector were reported in~\cite{jurgen:elba},
and publications from this year contain all details and performance
descriptions of this detector~\cite{jurgen:bigrichpaper,jurgen:israel}.

\begin{figure}[htb]
\begin{center}
\leavevmode
\epsfxsize=\hsize
\epsfbox{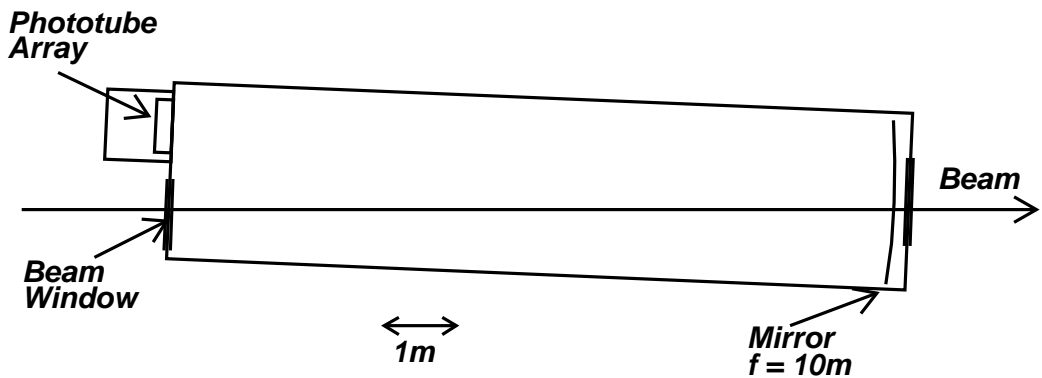}
\caption{Layout of the SELEX RICH detector.  The vessel has an overall
length of a little over $L=10\,\mbox{m}$, the mirrors have a radius of
$R=1985\,\mbox{m}$. The whole vessel is tilted by $2.4^\circ$ to avoid that
the particle trajectories go through the photomultipliers. Figure taken
from~\protect\cite{jurgen:bigrichpaper}.}
\label{jurgen:selexvessel}
\end{center}
\end{figure}

A layout of the vessel is shown in fig.~\ref{jurgen:selexvessel}.
The radiator gas is Neon at atmospheric pressure and room
temperature (see fig.~\ref{jurgen:neon}), filled into the vessel with a
nice gas-system~\cite{jurgen:gassystem}: 
First the vessel is flushed for about 1~day with 
${\rm CO}_2$ (a cheap gas). After this 
the gas (mostly ${\rm CO}_2$ and little air) is pumped in a closed system
over a cold trap running at liquid Nitrogen temperature, freezing out
${\rm CO}_2$ and the remaining water vapor.  At the same time Neon gets filled
into the vessel to keep the pressure constant.  This part of the procedure
takes about 1/2~day, and the vessel contains afterwards only Neon and 
about $100\,\mbox{ppm}$ of Oxygen which is removed by pumping the gas
over a filter of activated charcoal for a few hours, ending with an Oxygen
contents of $<10\,\mbox{ppm}$ in the radiator.  After this all valves were
closed and the vessels sits there for the whole data taking of 
more than 1~year at a slight ($\approx 1\,\mbox{psi}$)
overpressure\footnote{Actually the closed detector still sits untouched
in the PC4 pit at Fermilab.  We did not open it yet.}.

\looseness=1
The mirror array at the end of the vessel is made of $11\,\mbox{mm}$
low expansion glass, polished to an average radius of 
$R=(1982\pm 5)\,\mbox{cm}$, coated with Aluminum and a thin overcoating
of ${\rm MgF}_2$, which gives $>85\,\mbox{\%}$ reflectivity at 
$155\,\mbox{nm}$.
The quality of the mirrors was measured with the Ronchi
technique~\cite{jurgen:ronchi} to assure a sufficient surface quality of the
mirrors. The total mirror array covers $2\,\mbox{m} \times 1\,\mbox{m}$
and consists of 16~hexagonally shaped segments.  The mirrors are fixed with
a 3-point mount consisting of a double-differential screw and a ball bearing
to a low mass honeycomb panel.  The mirrors are mounted on one sphere, and
were aligned by sweeping a laser beam coming from the center of curvature
over the mirrors.

\begin{figure}[htb]
\begin{center}
\leavevmode
\epsfysize=8cm
\epsfbox{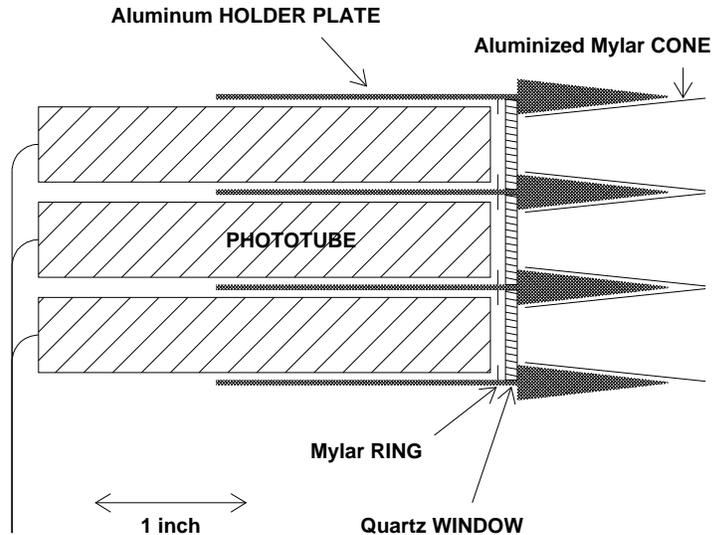}
\caption{Partial cross section through the Phototube Holder Plate.
For a more detailed description see text.
Figure taken from~\protect\cite{jurgen:bigrichpaper}.}
\label{jurgen:holderplate}
\end{center}
\end{figure}

The photo detector is a hexagonally closed packed $89\times 32$ array of
2848 half-inch photomultipliers. A side view is shown in 
fig.~\ref{jurgen:holderplate}.
In a $3\,\mbox{in.}$\ thick aluminum plate holes are drilled from both sides,
a $2\,\mbox{in.}$ deep straight hole holds the photomultiplier, and 
a conical hole on the radiator side holds aluminized mylar Winston cones, 
which form on the radiator side hexagons, leading to a total coverage of the
surface.  The 2848~holes are individually sealed with small quartz windows.
For the central region of the array, a mixture of Hamamatsu R760 and
FEU60 tubes were used, in the outside rows only FEU60 tubes are present.
The nearly 9000~cables (signal, hv, ground) are routed to the bottom (hv)
or top, where the signal cables are connected to 
preamp-discriminator-ecl-driver hybrid chips and finally readout via
standard latch modules\footnote{Since the phototubes are detecting single
photons, no ADCs are necessary.}.

\begin{figure}[htb]
\begin{center}
\leavevmode
\vspace{8cm}
\includegraphics{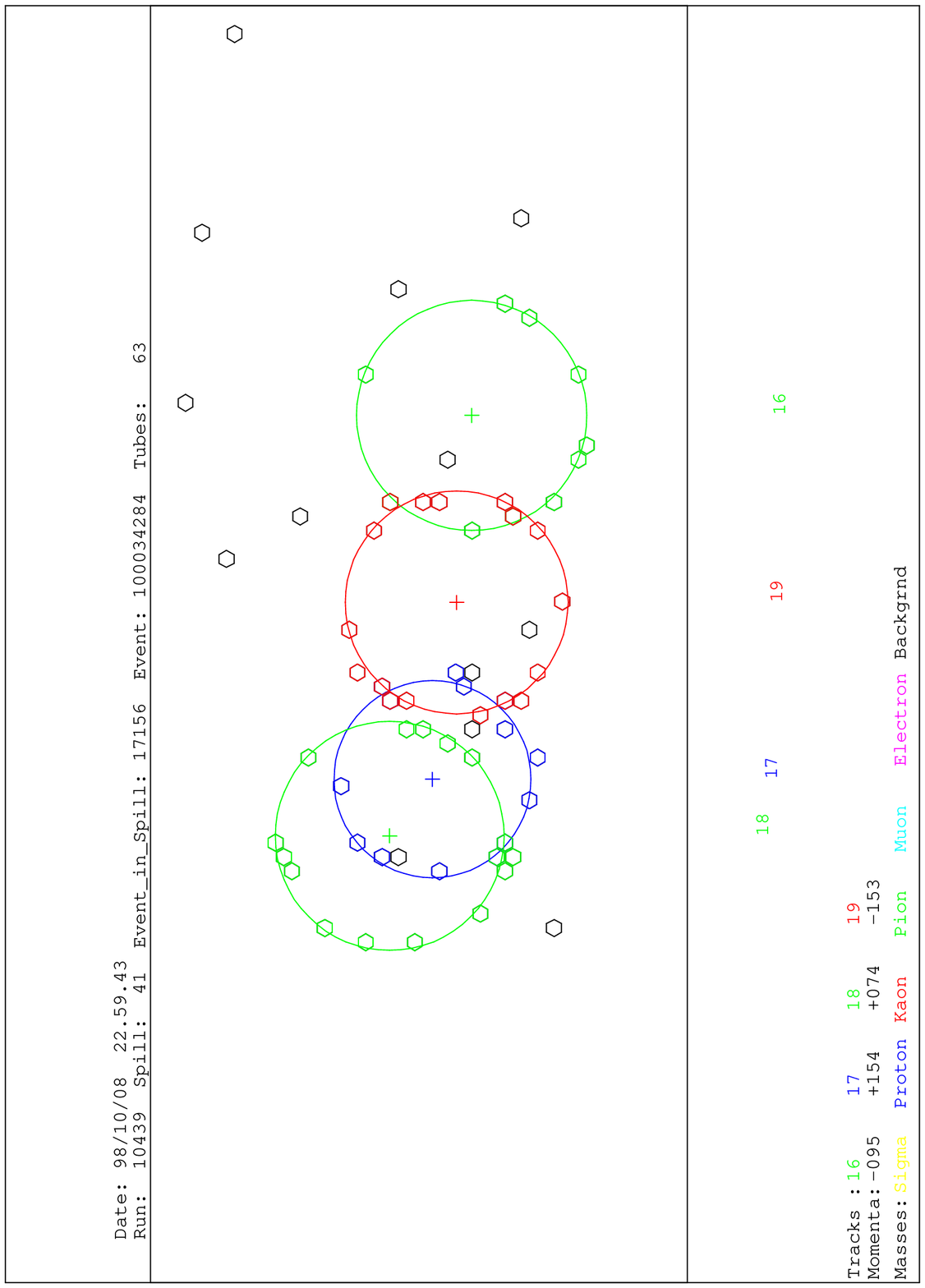}
\caption{Single event display. The small hexagons represent a hit phototube, 
the circle shows the ring for the most probable hypothesis, and the numbers
denote the track numbers, with there momenta shown at the bottom.}
\label{jurgen:singleevent}
\vspace{0.5cm}
\epsfxsize=\hsize
\epsfbox{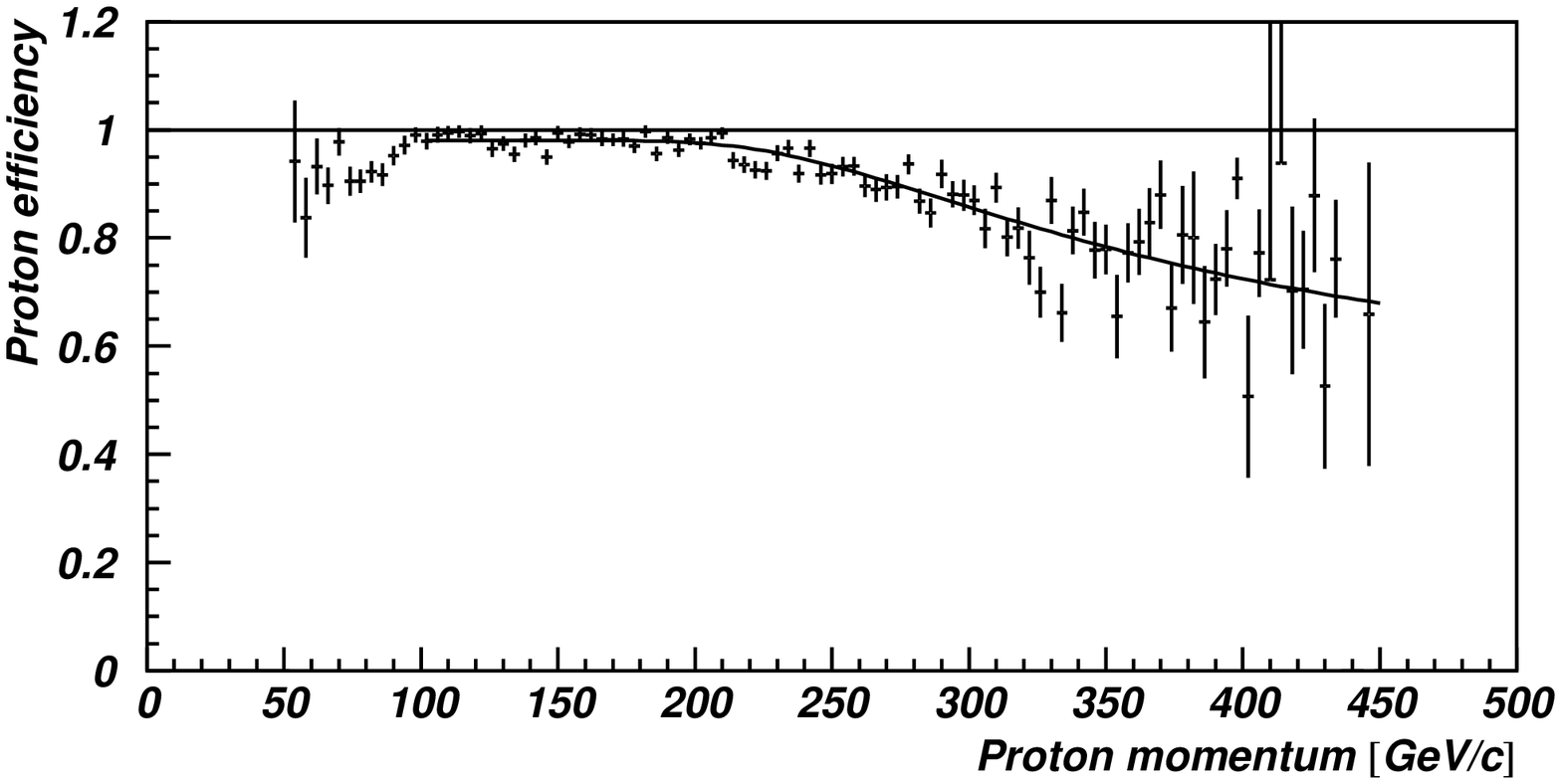}
\caption{Efficiency for identifying a proton as function of its momentum.
The likelihood of the track to be a proton has to be at least as big as to be
a pion.
Figure taken from~\protect\cite{jurgen:bigrichpaper}.}
\label{jurgen:proteff}
\end{center}
\end{figure}

A single event display of the detector is shown in 
fig.~\ref{jurgen:singleevent},
demonstrating the clear multitrack capability and the low noise of this
detector.  To analyze an event, the ring center is predicted via the known
track parameters, and a likelihood 
analysis~\cite{jurgen:likeli} for different hypothesis
(the momentum is known!)\ is performed to identify the particle. 
The final performance for this detector is shown in fig.~\ref{jurgen:proteff}.
The detector is nearly $100\,\mbox{\%}$ efficient, 
even below the proton threshold the efficiency is above $90\,\mbox{\%}$. 
In the SELEX offline analysis, the RICH is one of the first cuts applied
to extract physics results.  SELEX presented already several results 
at conferences~\cite{jurgen:selexa,jurgen:selexb,jurgen:selexc,%
jurgen:selexd,jurgen:selexe,jurgen:selexf}, and one paper is
submitted for publication~\cite{jurgen:selexg}.

\subsection*{Two RICHes for the CKM Experiment}

Last year a new experiment called CKM~\cite{jurgen:ckmproposal} was proposed
at Fermilab.
The goal of the experiment is to  measure the branching ratio for
$K^+ \to \pi^+ \nu\bar\nu$ to an accuracy of $10\,\mbox{\%}$ (SM prediction
is $10^{-10}$) to measure the CKM matrix element $V_{td}$.
To withstand the high expected physics background, the
experiment will use, in addition to a conventional magnetic spectrometer,
a velocity spectrometer consisting of two phototube RICH detectors, one
to measure the incoming $K^+$, the second the outgoing $\pi^+$.  The design
of the detectors is based on the SELEX RICH.  The HEP group in San Luis
Potos\'{\i} is involved in the design, construction, and testing of
parts of these detectors.

\clearpage
\section*{Acknowledgement}

The author wishes to thank the organizers for the opportunity to present
this course.  This work was partly financed by FAI-UASLP and CONACyT.
Figures~\ref{jurgen:selexvessel}, \ref{jurgen:holderplate}, and 
\ref{jurgen:proteff} 
are reprinted from Nuclear Instruments and Methods {\bf A431}, 
J.~Engelfried et al., The SELEX Phototube RICH Detector, pages 53-69, 1999,
with permission from Elsevier Science.

\end{document}